\documentclass[12pt,preprint]{aastex}

\slugcomment{}

\shorttitle{SBS0335-052E Super Star Clusters}
\shortauthors{Thompson et al.}

\begin{document}

\title{Super Star Clusters in SBS0335-052E}

\author{Rodger I. Thompson}
\affil{Steward Observatory, University of Arizona,
    Tucson, AZ 85721}
\email{rthompson@as.arizona.edu}

\author{Marc Sauvage}
\affil{Service d' Astrophysicque, CEA/DAPNIA, Centre d' Etudes de Saclay,}
\email{msauvage@cea.fr}

\author{Robert C. Kennicutt}
\affil{Institute of Astronomy, Cambridge University, Madingley Road, Cambridge,
CB3 0HA United Kingdom}
\email{robk@ast.cam.ac.uk}

\author{Charles Engelbracht}
\affil{Steward Observatory, University of Arizona,
    Tucson, AZ 85721}
\email{cengelbracht@as.arizona.edu}

\author{Leonardo Vanzi}
\affil{Pontificia Universidad Catolica de Chile, Department of Electrical
Engineering, Av. Vicu~na Mackenna 4860,Santiago, Chile}
\email{lvanzi@ing.puc.cl}

\author{Glenn Schneider}
\affil{Steward Observatory, University of Arizona,
    Tucson, AZ 85721}
\email{gschneider@as.arizona.edu}

\begin{abstract}

As one of the lowest metallicity star forming galaxies, with a nucleus of several 
super star clusters, SBS0335-052E is the subject of substantial current study. We
present new insights on this galaxy based on new and archival high spatial
resolution NICMOS and ACS images.  We provide new measurements and limits on the
size of several of the SSCs. The images have sufficient resolution to divide
the star formation into compact regions and newly discovered extended regions,
indicating a bi-modal form of star formation.  The star formation regions are
dated via the equivalent width of the Pa$\alpha$ emission and we find that two of 
the extended regions of star formation are less than 10 million years old.
Our previous finding that stellar winds confine the photo-ionizing flux
to small regions around individual stars is consistent with the new observations.
This may allow planet formation in what would traditionally be considered
a harsh environment and has implications for the number of planets
around globular cluster stars.  In addition the images pinpoint the regions 
of H$_2$ emission as located in, but not at the center of the two star forming 
super star clusters, S1 and S2.

\end{abstract}


\section{Introduction} \label{s-i}

SBS0335-052 is a Blue Compact Dwarf galaxy at a redshift of 0.0136,
\citep{izo97}, a distance of 54.3 Mpc and with an apparent V magnitude of 16.65 
\citep{thu96}. The close group of super star clusters (SSCs) in 
the galaxy's center has been and currently is the subject of intense
scrutiny. \cite{thm06}, hereinafter TSKEV, provides a recent summary of 
previous observations and investigations of the SSCs in SBS0335-052E.
Part of the attention is due to the extremely low metallicity of 
the galaxy.  The metallicity, as measured by 12 + log(O/H), 
in SBS0335-052E ranges between 7.20 and
7.31 \citep{pap06}.  SBS0335-052E has the fourth lowest metallicity of any 
star forming galaxy, surpassed only by its western component SBS0335-052W 
with the lowest measured metallicity,  J0956+2849, and the well known 
I Zw 18 \citep{izo07}. Because of its low metallicity and its now intense star 
formation, possibly due to an encounter with the nearby giant
spiral galaxy NGC 1376, it represents star formation and feedback 
in an environment very much like star formation in the early universe.  
SBS0335-052E is obviously in a very early stage of its star formation
since the observed stellar mass of $8 \times 10^7$ M$_{\sun}$ is only about
1/25 of the gas mass of $2.1 \times 10^9$ M$_{\sun}$
which is again small relative to the dynamical mass of $9 \times 10^9$
M$_{\sun}$ \citep{pus01}, most of which must be dark matter.  
A main conclusion of TSKEV was
that the majority of photo-ionizing radiation from the stars in
the actively star forming SSCs is absorbed within a few
stellar radii thus delaying the onset of negative feedback due
to photo-ionization. This conclusion is validated in the present 
study under the assumption that the diffuse Pa$\alpha$ flux around
the SSCs S1 and S2 is due to in-situ stars and not from ionizing
flux leaking out of the SSCs.

\begin{figure}
\plotone{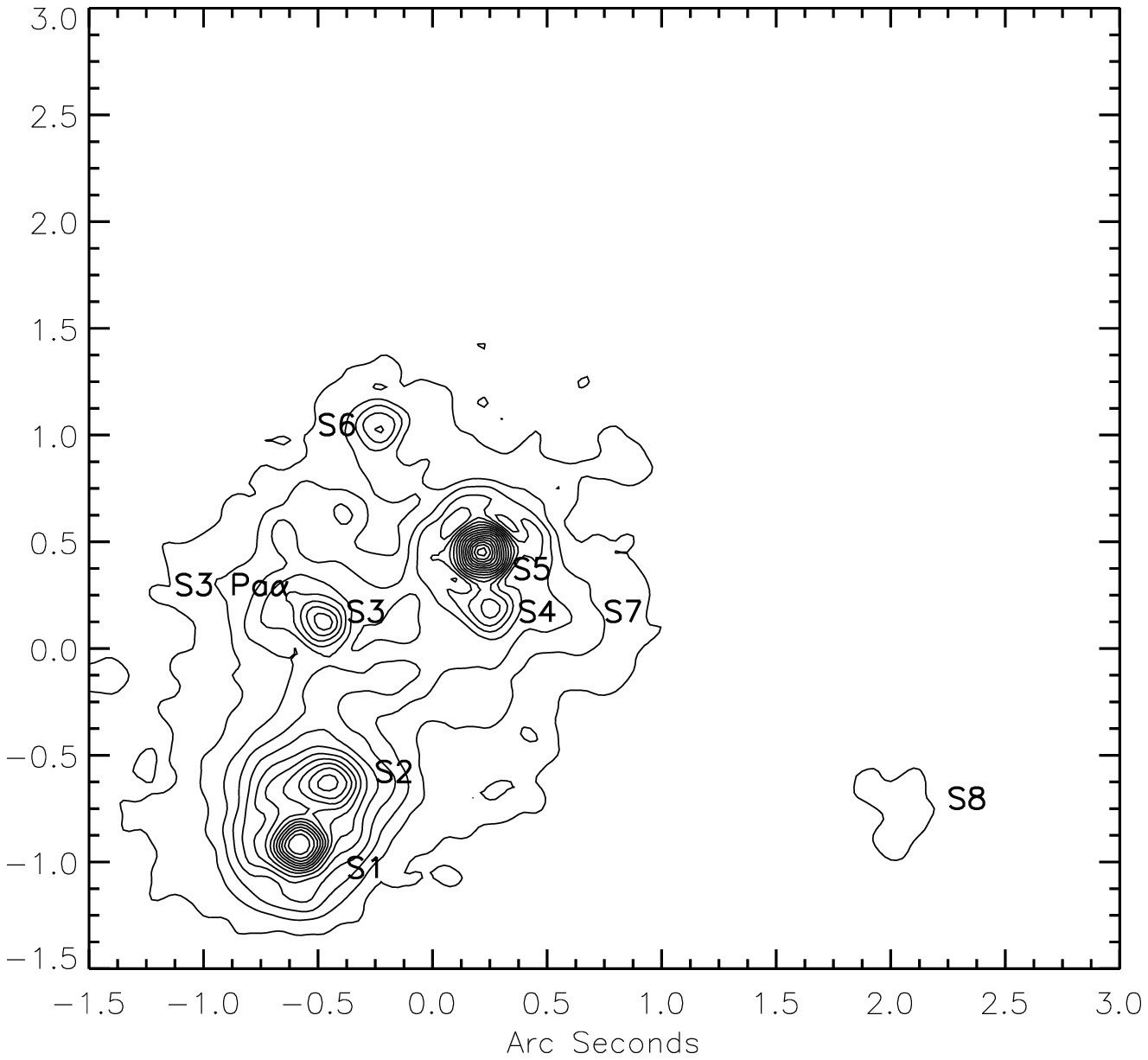}
\caption{This contour plot of the F160W (1.6 $\micron$) emission from in
SBS0335-052E shows the location of all of the sources described in this
paper.  The sources are displayed with a square root stretch to accommodate
the dynamic range of the source strengths.  The designation of sources S1-S6
are the the sources first named by \citet{thu97}. The sources S3-Pa$\alpha$,
S7 and S8 are named in this paper. S7 was originally marked as ? in
TSKEV. North is up and East is to the left.}
\label{fig-map}
\end{figure}

\begin{figure}
\epsscale{.8}
\plotone{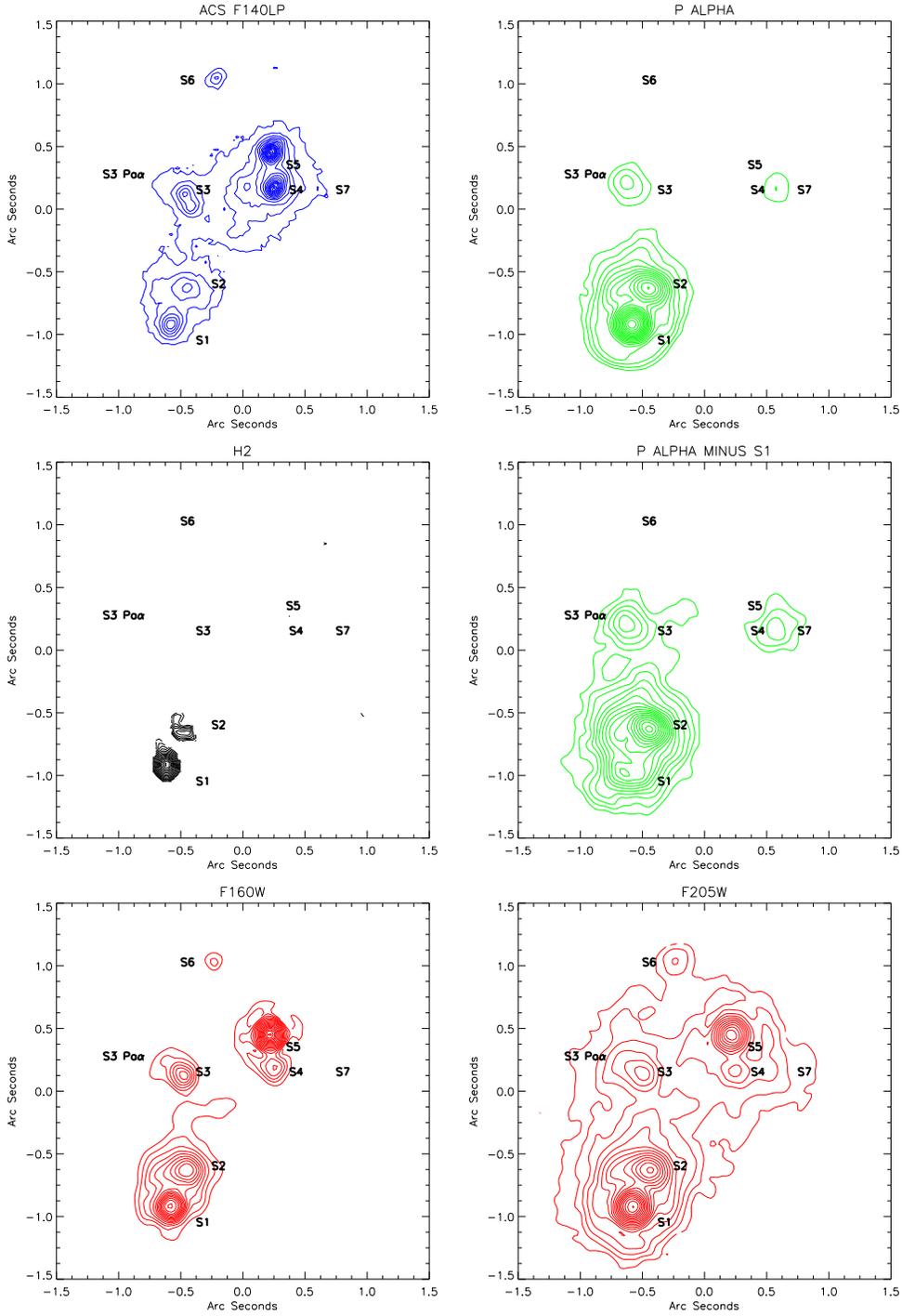}
\caption{Contour plots of ACS UV, Pa$\alpha$, H$_2$ and Near IR continuum
F160W and F205W images.  The plots have a square root stretch.  Contour lines
do not have the same flux levels between images. The contour plot titled P ALPHA
MINUS S1 is the Pa$\alpha$ image with the point-like source S1 subtracted to show the
underlying emission. North is up and East is to the left.}
\label{fig-mcon}
\epsscale{1.}
\end{figure}

The TSKEV study relied heavily on \emph{Hubble Space Telescope} Near Infrared 
Camera and Multi-Object Spectrometer (HST NICMOS) camera 3 observations of 
SBS0335-052E which  were part of the SINGS \citep{ken03} large survey
of nearby galaxies. The 0.2\arcsec pixel scale of camera 3 is not 
sufficient to resolve much of the detail of the feedback mechanism
on star formation operating in the clusters To get a higher spatial
resolution near infrared picture of the SSCs new observations with
the 0.075\arcsec pixel scale NICMOS camera 2 were taken in the
Pa$\alpha$ and H$_2$ filters.  The primary goal of this work is to
study in detail the feedback effect of the photoionizing flux generated
in the super star clusters on past and current star formation.  This study
also utilizes archival NICMOS camera 2 images in Pa$\alpha$ and the 
broad band filters F160W and F205W from observations taken by K.
Johnson (see \S~\ref{s-obs}).  Figure~\ref{fig-map} 
shows the location of all of the sources discussed
in this paper.  Sources S1-S6 are the sources originally designated 
from WFPC2 images in \citet{thu97}. The sources S3-Pa$\alpha$, S7 and
S8 are names ascribed by this work.  S3-Pa$\alpha$ is considered a
separate source from S3 as it is offset from the continuum source S3
and is most likely a completely separate source.  S7 is the
source that was designated by a question mark (?) in TSKEV and
S8 is diffuse source about $2.5\arcsec$ west of S1 and S2. Unfortunately
this source is marked as S7 in \citet{pap06} who did not have the
spatial resolution to resolve the source we designate as S7 as a 
separate source.  Unlike the NICMOS camera 3
images used in TSKEV the new, much higher resolution, NICMOS 
camera 2 images clearly resolve S1 and S2 and detect the separation
between S3 and S3-Pa$\alpha$.  Sources S3-Pa$\alpha$ and S7 only 
appear as slight extensions of the continuum emission in Figure~\ref{fig-map}
but are clearly shown in the Pa$\alpha$ contours in Figure~\ref{fig-mcon}.
The Advanced Camera for Surveys (ACS) image in Figure~\ref{fig-mcon} is 
an archival ACS solar blind channel (SBC) image in the UV F140LP filter 
\citep{kun03} which is also discussed in TSKEV.

\section{Observations} \label{s-obs}

The new observations presented in this paper were taken with the
HST NICMOS Camera 2 in 4 narrow band 
filters (F187N, F190N, F212N and F215N) centered at 1.87 $\micron$, 
1.90 $\micron$, 2.12 \micron\ and 2.15 \micron\ (Proposal ID 10856). 
These filters are designed to measure Pa$\alpha$ and the H$_2$ S(1) 
lines in the F187N and F212N filters respectively.  The redshift of 
SBS0335-052, however, shifts these lines to the F190N and 
F215N filters with the shorter wavelength filters acting as the 
continuum filters.  The program executed in 8 orbits.  Exposures 
were taken in all 4 filters in each orbit without moving HST during
an orbit. During each earth occultation the telescope was dithered one step 
in a spiral dither pattern with a dither step of $0.5\arcsec$. The
F190N filter which images the strong Pa$\alpha$ emission had an 
integration time of 320 seconds (Step32, NSamp=17) and the remaining
filters all had integration times of 768 seconds (Step64, NSamp=20) in
each orbit for a total of 2560 seconds for the F190N filter and 6144
seconds in each of the remaining filters.  The field of view of the
NICMOS camera 2 is $19.5\arcsec$ by $19.3\arcsec$ which easily encompasses
all of the known SSCs.

Additional archival images in the F160W, F205W, F187N and F190N filters 
(K. Johnson Proposal ID 10894), \citep{rei08}, were also retrieved. There are
8 images in the F160W and F205W filters.  The F160W images use STEP16
with 26 samples and the F205W images use STEP16 with 24 samples for
integration times of 304 and 272 seconds respectively.  There are 16
on target images in the F187N and F190N filters, again with STEP16 and
26 samples for F187N and 25 samples for F190N for integration times of
304 and 288 seconds. The total integration times of the archival images
are: F160W, 2432 seconds; F205W, 2176 seconds; F187N, 4864 seconds; 
F190N, 4600 seconds.

Archival ACS HRC and WFC1 images of SBS0335-052 from Proposal 10575
(PI Goran \"{O}stlin) in F220W, F330W, F550M, F435M filters and the FR656
ramp filter were also retrieved from the STScI MAST archive.  The 
combined drizzled images are used with no further processing.  These
images along with previous F122M and F140LP images from Proposal 9470 
(PI Daniel Kunth) have been used by \citet{ost08} to make H$\alpha$ 
and Ly$\alpha$ images. Table~\ref{tbl-obs} gives a summary of the 
observational parameters of the images used in this analysis.

Parallel to the NICMOS observations, images of the spiral galaxy 
NGC 1376 in 8 medium band filters were taken  with the ACS WFC. 
Two images in one of the 8 filters
were taken in each orbit. The ACS observations will be described in a 
separate publication. SBS0335-052 is probably engaged in a minor merger 
with NGC 1376 and the images were taken to see if there is any evidence
of the interaction.  No obvious signs of tidal disruption were seen in
the NGC 1376 images.  Given the small mass of SBS0335-052 relative to
NGC 1376 the absence of any disruption is not surprising. 

\section{Data Reduction} \label{s-dr}

All of the images were reduced with the NICMOS IDL based pipeline 
procedures operating on the raw frames as described in \citet{thm05}.  
The new images and the archival images were aligned separately since
they were at different orientations and had separate dither patterns.
The new images were aligned using offsets determined from the F190N image.  
The position of the point-like object
S1, described below, was determined by centroiding with the IDL procedure
IDP3 \citep{lyt99} for each image.  The same offsets were used for the 
other images in each orbit since HST was not moved during an orbit.  The
images in each filter were offset in IDP3 to the position of the images
in the first orbit and then median combined.  During processing visual 
inspection showed that two images, one line image and one continuum
image, were affected by a cosmic ray persistence
streak.  This occurs when there is a strong cosmic ray event before the 
integration begins and the image of the event persists in subsequent images.
The pixels involved in the event were identified and then not used in the
formation of the median image.  This means that the median is formed from
only 7 images rather than 8 for those pixels.
The archival images were aligned separately, using the image of S1 in each
filter as the fidelity of the pointing between images was not accurately 
known.

Subtraction of the associated narrow filter continuum image from the image in 
the emission line filter produces an emission line image.  The continuum images 
are scaled by the known responses of the line and continuum filters and detector 
QE and subtracted from the line image to produce a continuum free image. The 
spacing between the narrow line and continuum filters is only 0.03 microns
therefore the spectral slope of the SEDs, determined from the ground based 
spectra of \citet{van00}, has a negligible effect on the calculated line power. The 
Pa$\alpha$ image is the F190N image minus the scaled F187N image and the H$_2$ 
image is the F215N image minus the scaled F212N image.  The archival line emission 
images were reduced in the same manner and then rotated and 
shifted to match the new line images. A straight average was taken of the 
two images for the final line emission images.  The F187N image determines the
continuum level near the Pa$\alpha$ emission to measure the Pa$\alpha$ 
equivalent widths given in Table~\ref{tbl-ne2v}.

The archival wide filter continuum images were also rotated and shifted to match the
new images.  A median of all of the dithered F205W images was produced with
no shifts to produce a median sky image.  The median sky was then subtracted
from the F205W image to remove the thermal sky and instrument background. 
Contour plots of the ACS UV F140LP image from the data of \citet{kun04} and
contour plots of the new NICMOS continuum and line images are shown in 
Figure~\ref{fig-mcon}. Figure~\ref{fig-pa} shows the images of Pa$\alpha$, 
H$_2$, and the F160W continuum, along with ACS UV F140LP image. In 
these figures north is up in all cases.  Source S8 does not appear in
these figures due to the smaller size of the maps and images.

\section{Sizes of the point-like Sources, S1, S5 and S6} \label{s-size}

One aspect of SBS0335-052E, as pointed out earlier by \citet{har06},
 is that the majority of the star formation has 
occurred in very compact regions rather than being spread diffusely over 
the region as occurs in the Orion star forming complex in our own galaxy.  
Three of the sources S1, S5 and S6 display Airy rings indicative of a 
point-source.  We use NICMOS camera 2 observations of known single sources
observed to determine the NICMOS camera 2 PSF to get an estimate of the 
true size of these sources.  For the size of the Pa$\alpha$ emission we
use the PSF star GD-71 observed in the F190N filter and for the continuum we
use the PSF star HD 3670 observed in F160W filter. The PSF image of GD-71
is a median combination of 4 images at separate positions on the camera 2
detector array.  The images were shifted to a common position using IDP3. 
IDP3 uses  bi-cubic interpolation to resample and shift the images. The 
HD 3670 PSF 
image is the addition of two images again shifted and resampled in IDP3.
Each of these images were then resampled in an 8 times finer pixel scale,
again by bi-cubic interpolation in IDP3.  The images of S1, S5 and S6 were
then also resampled in the same way for comparison with the PSF star images.

The two resampled PSF star images were next convolved with 2D Gaussians with 
FWHM values of 3/8, 5/8, 3/4, and 15/16 of the \emph{original} camera 2 
0.075\arcsec pixel size.  The
radial profiles of the original PSF, the convolved PSF and the object images
at the 8 times finer scale were calculated in IDP3 and then plotted
in Figure~\ref{fig-size}. The convolved radial profiles were visually compared 
with the source profiles to determine the source size.  The comparisons were
only made in the inner $0.15\arcsec$ of the profile to avoid significant 
contamination from the diffuse background.  We did not attempt to subtract
the background since it is non-uniform with significant structure, particularly
for the Pa$\alpha$ background. 

The source S1 is resolved in both the continuum and in Pa$\alpha$ 
and is consistent with the radial profile of a Gaussian source with
a FWHM of $0.12\arcsec$ to $0.14\arcsec$ convolved with the NICMOS 
PSF.  At the 54.3 Mpc distance of SBS0335-052 \citep{thu97} the radius of 
S1, taken as 1/2 of the FWHM of the Gaussian, is between 16 and 18 pc. This 
is 4 to 5 times 
larger than the radius of S1 estimated in TSKEV based on the ACS F140LP
image.  That estimate was simply done by eye without a similar 
analysis using convolved PSFs and was clearly an underestimate.  This 
does not, however, alter any of the conclusions in TSKEV,
other than the speculation on the formation of intermediate mass black 
holes.  The S1 continuum source may be slightly more compact than
the Paschen $\alpha$ but the difference is less than the error in comparing
the observed profile with the broadened PSF profiles.  The known ``breathing''
of the HST secondary mirror produces temporal changes of the PSF that are
approximately plus and minus half the spacing between the PSF profile and the
first broadened profile.  The comparison may also be influenced by the 
extensive diffuse emission in the area.  The faint source S6 is also resolved 
with a size similar to S1 determined in the same manner as for S1.

Source S5 is not resolved in this image.  It is consistent with a
point source with an upper limit on the radius of 10 pc. It
does not have any Paschen $\alpha$ emission and is therefore probably
not undergoing intense star formation.  All of these sources have
sizes consistent with current day globular clusters and are most
likely the young predecessors of globular clusters.  

\section{The Individual Sources in SBS0335-052E}

The sources S1-S7 lie in a roughly north-south alignment with the
northern most source S6 separated from the southern most source
S1 by approximately $1.9\arcsec$ which is 490 pc at 54.3 Mpc.  A supernova cavity
lies about $4\arcsec$ (~1kpc) north of source S1.  There appears to
be a progression of ages of the clusters along the north-south
axis (Thuan, Izotov \& Lipovetsky 1997, TSKEV) starting with the supernova
cavity in the north to the very young sources S1 and S2 which
has lead to the speculation that the SSCs are the result of triggered
star formation propagating from the north to the south (TSKEV).
In the following discussion we have arbitrarily divided the sources
into sources with Pa$\alpha$ emission and those without.  We could
have equally well divided the sources into compact (SSCs) and 
extended.  In fact the compact sources without Pa$\alpha$ emission
are most likely older versions of the compact sources with Pa$\alpha$
emission.

\subsection{Sources With Pa$\alpha$ Emission} \label{s-pa}

Five sources, S1, S2, S3-Pa$\alpha$, S7 and S8 along with a diffuse background
have detectable Pa$\alpha$ emission and hence measurable star formation
rates. Unlike in TSKEV the high resolution NICMOS camera 2 images accurately
determine the distribution of Pa$\alpha$ flux between S1 and S2.  In addition 
to the Pa$\alpha$ flux associated with the continuum emission of S1 and S2, 
the images also show diffuse background Pa$\alpha$ emission in the region of 
S1 and S2 which was not evident in the previous lower spatial resolution 
observations.  The total diffuse Pa$\alpha$ flux is comparable to the sum of 
the fluxes from the sources and is due to either in-situ ionizing stars or to 
ionizing radiation that has leaked from the sources S1 and S2. The 
former is more likely than the latter (See \S~\ref{ss-ds1ds2}). If the
emission is due to ionization by in-situ stars it is not clear whether it 
is due to a diffuse distribution of individual stars or a distribution of 
smaller clusters that is not resolved in the present observations. 

Two types of photometry were carried out on the sources.  The first is
aperture photometry with a 5 pixel diameter (0.378\arcsec) that encompasses
all of the flux inside the first Airy ring of a point source.  This is
$74\%$ of the total flux of a point source in the F160W filter and $69\%$
in the F190N filter.  Larger apertures begin to infringe on adjacent sources.
To first order adjacent sources add approximately the same flux to the aperture
as falls outside the aperture from the primary sources.  We therefore do 
not correct the aperture flux for the percentage of primary object flux in
the aperture.  The background was measured in annular rings 3 pixels thick 
around the sources.  The inner and outer diameters of the annular rings were 
adjusted to avoid flux from adjacent sources. The measured backgrounds, 
adjusted to the area of the photometric aperture were subtracted from the 
source flux. The photometric error is taken as $10\%$ of the total flux
for the bright compact sources and is dominated by the systematics of source
overlap and background subtraction. 

The equivalent widths were similarly measured in a 6 pixel (0.454\arcsec)
diameter aperture. The continuum measure for the equivalent widths is 
provided by the narrow band F187N filter which is adjacent to the F190N
band that measures the Pa$\alpha$ flux.  This filter accurately measures 
the continuum level at the location of the Pa$\alpha$ line and is free
of other contaminating emission lines. The larger aperture was used to 
increase the signal to noise of the measurements in the F187N filter which
has much lower flux than the Pa$\alpha$ filter or the broad band F160W
filter.  The smaller aperture photometry agreed
within their signal to noise ratios with the larger aperture measurements.
Table~\ref{tbl-flux} gives the line and continuum fluxes for all of the
sources along with the equivalent widths of the Pa$\alpha$ line.  It is
evident that the equivalent widths of the extended sources S7 and S8 are 
significantly higher than for the compact sources S1 and S2.  Again note that S7
is the source denoted by a question mark in TSKEV and S8 is the source
marked S7 in \citet{pap06}.  Unlike in TSKEV the new observations are deep enough
to detect the faint continuum associated with the extended sources and are
of high enough resolution to discern the offset between the continuum and
line emission in S3.  In the following we discuss the nature of the 
sources.

\subsubsection{Sources S1 and S2} \label{ss-s1s2}

The Pa$\alpha$ emission in S1 and S2 is coincident with the continuum
emission.  In the new images it is clear that the Pa$\alpha$ emission 
in the region of S1 and S2 has at least three components, S1, S2 and diffuse, 
therefore, the extinction of A$_V$ = 12.1 derived from the comparison 
of Pa$\alpha$ and Br$\alpha$ flux ratios in TSKEV is an average extinction. 
The extinction curves of \citet{rie85} were used to compute the extinction.
The extinction for any of 
the individual components may be different than the average.  If, however,
the extinction is primarily foreground then the extinction effects all
sources equally and the derivation is valid.  We note that the 1.6 \micron\ 
to 2.05 \micron\ color is similar for S1 and S2 indicating that they may have 
the same extinction.  The color of the diffuse continuum emission is blue 
rather than red, however, which supports the idea of less extinction in the 
regions of diffuse emission.

To get an estimate of the spatial distribution of the extinction we 
produced a H$\alpha$ image 
from the archive ACS HST images of SBS0335-052 from Proposal 10575 
(PI G. \"{O}stlin) by subtracting the F550M image from the FR675N ramp
filter image.  The F550M image was multiplied by a factor of 0.2 to 
account for the differences in throughput. The factor was determined
by the requirement that the sources with no Pa$\alpha$ emission, and
therefore presumed to have no H$\alpha$ emission, subtract to zero.
This does not guarantee an accurate subtraction in the sources with
hydrogen line emission since they are generally redder than the 
non-emission line sources. The H$\alpha$ image was then resampled to the
spatial resolution of the NICMOS Pa$\alpha$ image.  The ratio of the two 
images is an indicator of the relative extinction and is 
scaled to unity at the location of S1. The differences in wavelength
which changes the diameter of the Airy rings limits the accuracy of 
this procedure. The extinctions of S1 and S2 are equal within the
signal to noise of the image. The extinction of the extended sources 
S3-Pa$\alpha$, S7 and S8 are also equal to S1 and S2 within the rather 
low signal to noise.  The extinction in the diffuse region is 1/2 of 
that of S1 and S2 in terms of absolute flux. We therefore decrease the 
extinction in the diffuse flux by a factor of two in computing the properties 
of the diffuse emission.  The extinction map shows significant structure
due to the differences in spatial resolution and we estimate the 
accuracy of the relative extinctions derived from the map to be $20\%$. 

The volume emission measure (${N_e}^2V$) for S1 of $7.9 \times 10^{65}$
cm$^{-3}$ is almost exactly the same as the previous measurement in TSKEV of
$7.2 \times 10^{65}$, therefore the primary conclusions on the nature
of S1 remain valid.  Note that throughout the discussion of the Pa$\alpha$
sources the ionized gas temperature is assumed to be 20,000K as discussed
in TSKEV.  S1 has a stellar mass of $1.5 \times 10^7$ M$_{\sun}$, see 
\S~\ref{s-mass}, with stars of less than 1 M$_{\sun}$ probably not yet on the main
sequence.  There are 8650 stars earlier than O9 and the luminosity of
the stellar population is $3.4 \times 10^9$ L$_{\sun}$ based on the
output of Starburst99 \citep{lei99} if we only account for the Pa$\alpha$ 
flux in the point-like object. A further discussion of the Starburst99
calculation is given in \S~\ref{s-age}. The star formation rate
is 1.2 M$_{\sun}$ per year based on a recombination Case B recombination
rate for 20,000K \citep{ost89} and the SFR to Q(H$^0$) ratio given
by \citet{ken98}. The specific star formation rate, star formation rate
divided by the mass of the S1, of $1.4 \times 10^{-7}$ yr$^{-1}$  is 
relatively high. The specific star formation rate for the entire galaxy of 
$3.7 \times 10^{-9}$ yr$^{-1}$, however, is quite low.

We now resolve S2 from S1 and obtain a volume emission measure of 
$5.9 \times 10^{65}$ cm$^{-3}$ for S2, somewhat less than the arbitrary 
half of the total S1 plus S2 emission measure that was assigned to S2 
in TSKEV.  The parameters discussed above for S1 simply 
scale for S2 by the ratio of the volume emission measures.  The difference
between the values of the emission measure for S1 plus S2 in this work
versus TSKEV is due to the separation of the sources from the diffuse
background, some of which was included in the aperture measurement of
Pa$\alpha$ in the previous work. Unlike S1, S2 is clearly not a point 
like object. It is irregular in shape and extended in the East-West
direction.  Its average spatial extent is $0.05\arcsec$ larger than
S1 which puts its radius at 30 pc averaged over the East-West and 
North-South directions. 

\subsubsection{Diffuse Emission Around S1 and S2} \label{ss-ds1ds2}

The diffuse emission is the Pa$\alpha$ emission that remains after
all of the emission associated with the sources S1-S8 is subtracted.  The 
associated diffuse continuum emission is defined in the same way.  The 
majority of the diffuse emission, defined in this manner, is in the region 
around S1 and S2.  In the commonly used radio emission measure the volume 
emission measure of $5.5 \times 10^{65}$ cm$^{-3}$ is equivalent to that of 
$1.2 \times 10^4$ O7 stars. The volume emission measure of the diffuse 
emission is more than 1/3 of the emission measures of S1 and
S2.  This implies that the current star formation is roughly equally 
spread between diffuse and compact regions. 
We consider two possible interpretations of the diffuse Pa$\alpha$
emission.  The first is ionizing radiation from S1 and S2 that escapes the
clusters and ionizes the surrounding gas.  The second is an underlying
population of young stars.

The volume emission measure of the diffuse gas is $40\%$ of the volume emission
measure of S1 and S2. The diffuse emission therefore requires some source
of ionizing flux at the level of $40\%$ of that produced by S1 and S2.  This
assumes that all of the ionizing photons from the source of ionizing flux
for the diffuse emission are absorbed by the diffuse gas and none escape
from the region. If the source of ionizing photons is photons that escape
S1 and S2 but are trapped by the diffuse gas the mass and luminosity of the 
two sources must increase by a similar percentage.  This would place the mass 
of S1 and S2 at $2.1 \times 10^7$ M$_{\sun}$ and 
$1.6 \times 10^7$ M$_{\sun}$ respectively.  The luminosities for S1 and S2
become $4.8 \times 10^9$ L$_{\sun}$ and $3.5 \times 10^9$ L$_{\sun}$. These
are still reasonable parameters for the 2 SSCs but they lie at the high
side of known clusters, eg. \citep{zha99,dow08}. A difficulty may 
be that this makes the percentage of dust re-emitted luminosity of
$1.2 \times 10^8$ L$_{\sun}$ \citep{hou04,eng08}, only $20\%$ of the total 
luminosity of S1 and S2.  The discussion of the H$_2$ emission in
\S~\ref{s-h2}, however, shows that this is a reasonable number.  Another
difficulty could arise if the assumption that all of the ionizing flux
produces observable Pa$\alpha$ emission is false.  If a large fraction of
the ionizing flux escapes the region then the mass and luminosity requirement
can become unrealistic.  Also if the diffuse emission is simply ionized gas 
with no stellar continuum its Pa$\alpha$ equivalent width should be higher
than that of S1 and S2, instead it is lower. The equivalent width calculations
described in \S~\ref{s-age} in fact show that the diffuse emission has the
lowest equivalent width of any Pa$\alpha$ source. This is very strong evidence
against the concept that the diffuse Pa$\alpha$ is due to UV leakage from
the clusters.  The discussion in \S~\ref{s-age} also indicates that the 
diffuse emission stellar population is most likely older than the other
Pa$\alpha$ sources.

The possibility that the diffuse ionization region is produced by an in-situ
diffuse distribution of stars is a more plausible explanation.  Accounting for
the diffuse emission with in-situ stars requires a stellar mass of $40\%$ 
of the combined masses of S1 and S2 if the stars are at the same epoch
of evolution as the two SSCs.   The net star formation rate of the diffuse
component in this 
explanation is 0.85 M$_{\sun}$ yr$^{-1}$.  If this is the correct explanation
for the diffuse emission then star formation in SBS0335-052 is occurring
in a bi-modal fashion.  The majority of the stellar mass is in the 
form of compact objects, S1-S6, but the current star formation is roughly
divided 2 to 1 between the compact and diffuse regions whereas current star 
formation in the Milky Way appears to be predominantly in diffuse regions. 

\subsubsection{Source S3-Pa$\alpha$} \label{ss-s3pa}

S3-Pa$\alpha$ is a source with a Pa$\alpha$ equivalent width very close to
those of S1 and S2.  To delineate the Pa$\alpha$ emitting region from the 
continuum source S3 we have designated the emission region S3-Pa$\alpha$. 
The separation of S3-Pa$\alpha$ is the reason that \citet{pap06} found
an offset between the continuum and the line emission near S3. 
As with the diffuse emission the question is whether this is a separate star 
forming source or is it gas that is ionized by the double source S3 or 
possibly the more distant sources S1 and S2.  S3-Pa$\alpha$ has a volume 
emission measure of $2.8 \times 10^{64}$ cm$^{-3}$ which is equal to 
that of $2.9 \times 10^3$ O7 stars. If this volume emission measure is due to sources 
other than in-situ stars it would have to be due to intercepted ionizing 
flux from one of the other sources.

The lack of detectable Pa$\alpha$ flux at the location of S3 argues
against it as the source of the ionizing flux.  The only possibility
would be if the the region around and in S3 is largely free of gas.  
The observed Pa$\alpha$ region subtends 2.2 steradians as seen from
the location of S3 if its radial dimension is the same as its extent
on the sky and therefore receives approximately $17\%$ of the 
emitted flux from S3 if it is emitted isotropically.  Note that as in 
the discussion of the diffuse emission we assume that the source of
UV flux, the stars, emit radiation isotropically.  Sources can be
``beamed'' by large scale distributions of dust but dust structures
simply block ionizing flux, they do not enhance the flux along any
particular line of sight.  The volume emission measure for S3 would 
then have to be $7.6 \times 10^{65}$ cm$^{-3}$, almost as high as S1.   
The requirements for ionization by S1 and S2 are even more formidable.  
S3-Pa$\alpha$ subtends $3.4 \times 10^{-2}$ steradians
as seen from S1.  The ionizing flux from S1 and S2 would then have to
be 370 times larger than listed in Table~\ref{tbl-ne2v}, increasing 
their stellar mass to almost $10^{10}$ M$_{\sun}$.  This is clearly
not possible. 

The most likely source of ionization of S3-Pa$\alpha$ is in-situ young
stars.  This is the explanation that requires the minimum amount of 
ionizing flux as is the case for the diffuse flux discussed earlier.
The star formation in this source is then in a more
extended region than in the SSCs, again showing the bi-modal nature of
high pressure, low metallicity star formation.  Another argument in favor of local
star formation in S3-Pa$\alpha$ is that the equivalent width of the
Pa$\alpha$ flux is almost exactly the same as in S1 and S2.  This 
argues that S3-Pa$\alpha$ is the location of early star formation
such as in S1 and S2 while the diffuse emission around S1 and S2, 
which has a lower equivalent width, is due to an older stage of
star formation as is further discussed in \S~\ref{s-age}.

\subsubsection{Sources S7 and S8} \label{ss-s7s8}

Two of the Pa$\alpha$ emitting sources are distinguished by significantly
larger equivalent widths than the rest of the sources.  They are
S7, the previously known source that was marked with a question mark in
TSKEV and S8, $2.59\arcsec$ (664 pc) from S1 as shown in 
Figure~\ref{fig-map}.  Note that both S7 and S8 are clearly
visible in the images presented by \citet{ost08}, eg. their Figure 3,
but were not commented on in the paper.  Both of these sources are extended
and far away from any other known ionizing sources, particularly for S8.
It appears clear that these must be sites of new star formation and the 
ionization is due to in-situ stars.  Their volume emission measures, as
shown in Table~\ref{tbl-ne2v}, are about a factor of 10 less than S1 and
S2 and the total mass of stars is less.  The star formation rates
are 0.14 and 0.043 M$_{\sun}$ yr$^{-1}$ respectively.  The extinction
map discussed in \S~\ref{ss-s1s2} gives the same average extinction in
S7 and S8 as for S1 and S2, but, the low signal to noise of the continuum
make variations by a factor of 2 quite possible.

\subsection{Sources S3, S4, S5 and S6} \label{s-s3456}

\begin{figure}
\epsscale{0.8}
\plotone{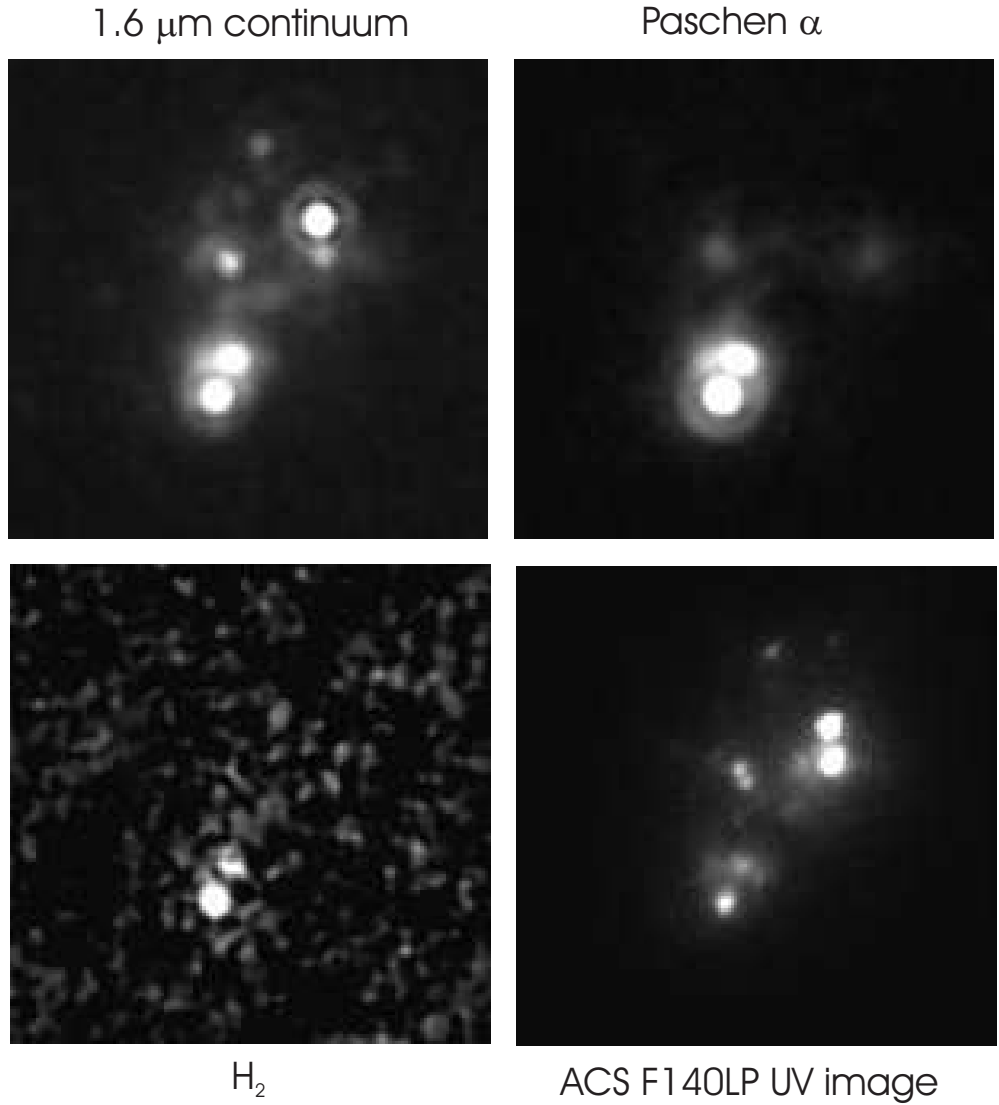}
\caption{Linear stretch 40 by 40 pixel NICMOS camera 2 images of SBS0335-052
in the continuum, Pa$\alpha$, H$_2$ and the ACS F140LP UV image.  The images are 
approximately $3\arcsec$ on a side. Note the Airy rings around S1, S5 and S6
(faint) in the 1.6 \micron\ continuum image and around S1 in the Pa$\alpha$
image.  North is up and East is to the left.} 
\label{fig-pa}
\epsscale{1.}
\end{figure}

\begin{figure}
\plotone{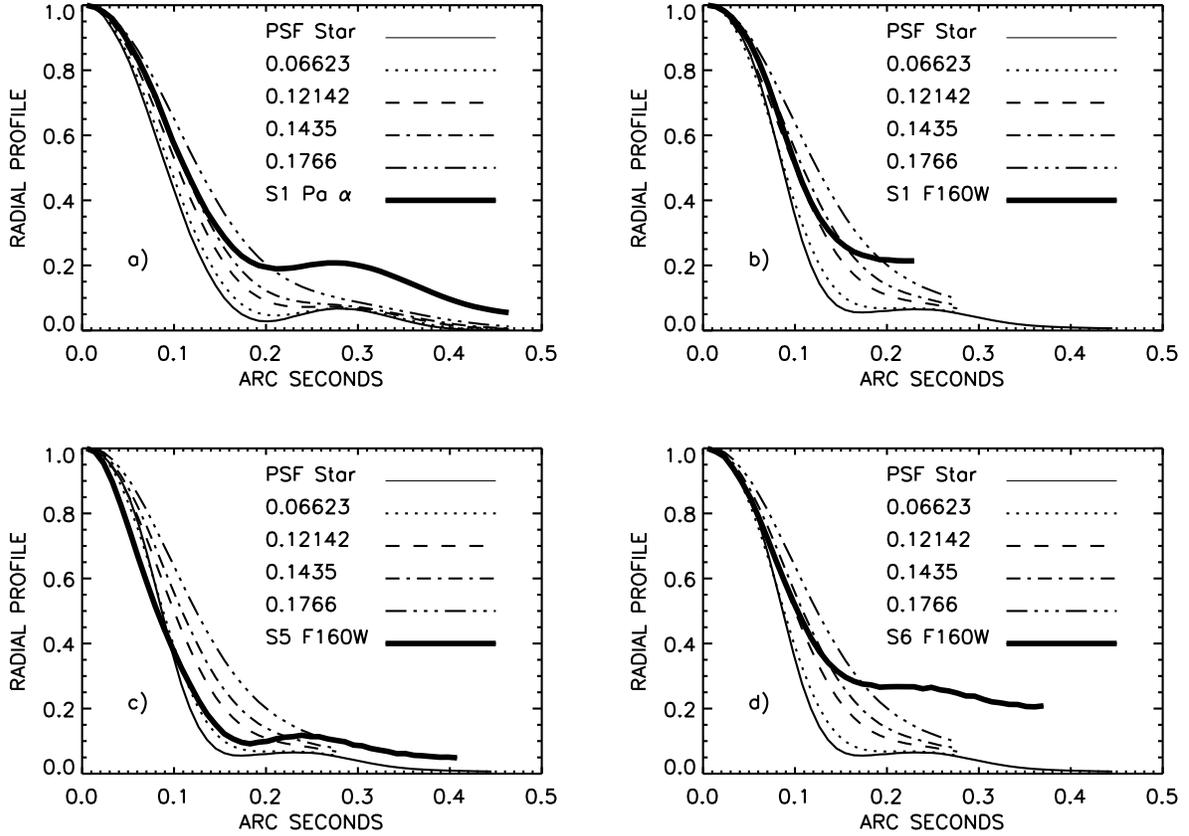}
\caption{This figure plots the radial profile of a) the Paschen $\alpha$ emission
(F190N filter) from
source S1 and the F160W continuum profiles of b) S1, c) S5 and d) S6 compared to the 
radial profiles of a standard PSF star and that star convolved with Gaussian profiles 
with the FWHM values listed in the legend.  The values correspond to 3/8, 5/8, 3/4 and
15/16 of a pixel respectively.  The profiles do not correspond to the PSF profiles in
the regions beyond $0.15\arcsec$ due to diffuse emission or adjacent sources. See the 
text for a discussion of the source sizes. The names of the sources appear in the 
last row of the in figure legends.
\label{fig-size}}
\epsscale{1.}
\end{figure}

The sources S3, S4, S5 and S6 are distinguished by the absence of any associated
Pa$\alpha$ emission indicating a lack of ionized gas and current star 
formation in these sources.  Their continuum 1.6 to 2.05 \micron\ colors are bluer than
S1 and S2 which is most likely a function of the foreground extinction
rather than their intrinsic color.  Source S3 deserves special mention 
as it may be the most enigmatic of all of the non-Pa$\alpha$ sources. 
In the ACS UV F140LP image in Figure~\ref{fig-pa} S3 
is clearly double, also seen in Figure 1 right of TSKEV, with a spacing 
of $0.08\arcsec$ (23 pc) in the northeast-southwest direction on the sky.  
The infrared continuum is centered on the northeast component with no 
indication of a southwest source or extension. The continuum fluxes for 
S3 are also hard to explain.  At UV wavelengths the two components of S3 
are roughly equal in brightness (Figure~\ref{fig-pa} ACS UV F140LP) while 
at infrared wavelengths (Figure~\ref{fig-pa} 1.6\micron\ continuum) only 
the northeast component is seen but 
it is quite bright. A possibility is that the northeast component has an
intrinsic brightness much brighter than the southwest but is much more heavily
obscured so that at UV wavelengths they appear to be a roughly the
same brightness.  This explanation is possible but appears to be a
bit contrived. The obscuring dust would have to be a significant distance 
in front of the northeast component of S3 otherwise we should see mid and 
far infrared emission from that location. On the other hand, it must be
small enough to only obscure the northeast source.

The discussion in \S~\ref{s-age} indicates that the non-Pa$\alpha$ sources
are older than the other sources which is consistent with triggered star 
formation starting in the north and propagating southward.  Only S3 is 
inconsistent with this progression since it is very near the young star 
forming region S3-Pa$\alpha$.  S5 is somewhat unique in that it is extremely 
compact as discussed in \S~\ref{s-size}. S5 could possibly be a young 
globular cluster that has undergone core collapse. All of the non-Pa$\alpha$
sources have probably had most of the residual gas swept out of them and are 
bound young globular clusters.

\section{The ages of the sources} \label{s-age}

Table~\ref{tbl-flux} gives equivalent widths for the Pa$\alpha$ emission
sources.  The equivalent width is relative to the continuum measured
in the F187N narrow line filter which is adjacent in wavelength to the
F190N filter that measures the Pa$\alpha$ flux.  Use of the F160W or 
F205W filter images would give higher signal to noise but the uncertainty
in the detailed spectrum and the inclusion of other emission lines in the
broad filter compromises the accuracy of the equivalent width.  In an
attempt to determine the ages of the emission line sources we ran a
Starburst99 \citep{lei99} simulation with a stellar mass range of 
0.1-120 M$_{\sun}$, a standard Kroupa IMF, a metallicity of 0.001,
time steps of $10^5$ years and an instantaneous burst of star formation.  
The luminosity of the Pa$\alpha$ line was
calculated from the output Pa$\beta$ luminosity using Case B recombination
and a temperature of 20,000K. From this line luminosity and the computed continuum
flux at the wavelength of Pa$\alpha$ the equivalent width of Pa$\alpha$ was
calculated and plotted versus age in Figure~\ref{fig-eqw}. Note that the 
continuum includes both stellar and nebular emission.  The plot is
marked with asterisks at the equivalent width appropriate for each source.
The position marked for S8 is at the oldest consistent time.  This equivalent width
is achieved at two earlier times but the time spent at those values is
very brief, $10^5$ and $1.5 \times 10^5$ years, as opposed to $1.3 \times 10^6$
years at the marked position.  The dip in equivalent width around 
$4 \times 10^6$ years is due to a rise in the infrared continuum rather 
than a dip in the Pa$\alpha$ luminosity.  

The parameters of the Starburst99 simulation have little effect
on the age sequence and ages of the sources for an instantaneous starburst.
It is only the lifetimes of the high mass stars that affect the evolution
of the ionizing flux.  Higher metallicities would decrease the ionizing
flux per unit mass and different IMFs would change the total mass needed
to achieve the observed ionizing fluxes but not the time evolution.  To 
check this assumption we produced another Starburst99 simulation with a
Salpeter rather than Kroupa IMF.  The resultant equivalent to Figure~\ref{fig-eqw}
was indistinguishable from Figure~\ref{fig-eqw}.  These changes only
 affect the mass estimate discussed in \S~\ref{s-mass}.  If, however, the 
star formation in each of the clusters is not short relative to the present
age of the cluster the ages could be quite different from those derived
here.  The ages from Figure~\ref{fig-eqw} are only valid if the star formation
epoch is short relative to the age of the clusters.

Figure~\ref{fig-eqw} indicates that S7 and S8 are young star forming regions
with ages less than 10 million years with S7, the youngest, at about 4 million
years.  S1, S2 and S3-Pa$\alpha$ all formed at roughly the same epoch and are now about
10 million years old.  The diffuse emission is from an older group of stars
formed a million years before the S1, S2, S3-Pa$\alpha$ group.  The sources
without detectable Pa$\alpha$ emission are at least 15 million years old
or older.  This shows a general progression in age of the SSCs with oldest
sources in the north and the youngest in the south, evidence for triggered
star formation running from north to south.  New star formation is now
occurring in S7 and the more distant S8.  A 1 km/sec wind would propagate 10
pc in 10 million years which makes it doubtful that S8, which is 340 pc from
S1 and S2, is causally affected by the SSCs.

\section{Stellar Masses} \label{s-mass}

The Starburst99 output provides an estimate of the stellar mass for those objects
with Pa$\alpha$ emission.  The Starburst99 calculation used a stellar mass of
$10^8 M_{\sun}$, therefore, the ratio of the observed Pa$\alpha$ power versus
the calculated Pa$\alpha$ power times $10^8 M_{\sun}$ at the age given in
Figure~\ref{fig-eqw} gives the mass.  The masses are given in the last column of 
Table~\ref{tbl-ne2v}.  As discussed in \S~\ref{s-age} the masses are sensitive
to the parameters used in the Starburst99 model, particularly on the IMF 
and metallicity
used.  The metallicity is appropriate for that measured in the clusters but we
do not have sufficient information to accurately determine the actual IMF and
many of the low mass stars are still probably in their pre-main sequence evolutionary
stages. The largest mass components are the two compact sources S1
and S2 plus the diffuse component which total to a mass of $4.65 \times 10^7 M_{\sun}$. 
If the SSCs with no Pa$\alpha$ are included with an average mass of $10^7 M_{\sun}$ 
per SSC then the total mass is on the order of $8 \times 10^7 M_{\sun}$.  Taking the 
total gas mass as $2.1 \times 10^9 M_{\sun}$ \citep{pus01} then the ratio of 
total stellar mass in the nucleus of SBS0335-052E to the gas mass of 1 to 26 
is significantly higher than the average nuclear cluster mass to gas ratio 
found by \citet{set08}. The average nuclear cluster mass to gas ratio in 
\citet{set08}, however, is dominated by nuclear clusters in large spiral 
galaxies rather than in BCD low metallicity galaxies.
The objects S3-Pa$\alpha$, S7 and S8, which have less concentrated star formation
have significantly less mass, with S7 having only $9.4 \times 10^4 M_{\sun}$.  This
may represent star formation in what we would term open clusters and they may not
remain bound for very long.

\section{H$_2$ Emission in S1 and S2} \label{s-h2}

A new finding of this investigation is the precise location of the
H$_2$ emission in S1 and S2.  Previous ground based observations by
\citet{van00} did not have the spatial resolution to resolve the 
sources.  The contour map in Figure~\ref{fig-h2} shows separate emission 
regions in both S1 and S2.  The emission region in S1 is consistent
with being point-like while the emission in S2 is extended as is the 
continuum emission.  Both of the regions are offset from the center
of the continuum emission.  The H$_2$ emission region in S1 is offset
by $0.04\arcsec$ (11.6 pc) to the east from the continuum center and 
the H$_2$ region in S2 is offset by the same amount to the south.  
These offsets are less than the radii of the clusters so that the 
molecular emission regions exist inside the region of Pa$\alpha$
emission. There is no chance that the offsets are due to misalignment
of the images as the new H$_2$ and Pa$\alpha$ images were taken in
the same orbit for each dither position with no change is HST attitude.
Taken alone these images clearly show the offset.  The additional 
archival Pa$\alpha$ images were then registered to the new Pa$\alpha$
images and the same offset was present.  It is remotely possible that
the two H$_2$ regions are located along the line of sight to S1 and S2
but outside of the clusters.  This might be considered a possibility
if only one SSC coincided with a H$_2$ source but it is highly unlikely
that two SSCs lie along the line of sight to the only two H$_2$
sources. 

The survival of molecular gas inside the sources is another
line of evidence that the photo-ionizing flux does not reach throughout
the clusters.  The molecular regions are also the most likely sources
of the observed mid-infrared emission found by \citet{hou04} and
further discussed by \citet{eng08}.  If these are the emission regions
and they are small dust regions embedded in the cluster then they are
not capable of absorbing all of the luminosity of the cluster and 
re-emitting it in the infrared.  This accounts for the low IR luminosity
to total luminosity ratio discussed in \S~\ref{ss-ds1ds2}.  Their
location inside the cluster also probably accounts for the high
temperature of the observed dust emission.

The calculation of the mass of H$_2$ needed to produce the observed
emission depends on the temperature of the H$_2$ gas and the ratio
of thermal to fluorescent emission, neither of which is accurately
available.  \citet{van00} find that the emission is a mixture of
fluorescent and thermal emission which makes the line ratios an
unreliable monitor of the thermal temperature.  To get a rough 
estimate of the amount of H$_2$ needed to produce the observed
(1-0) S(1) emission we 
calculate the H$_2$ abundance needed for a 2500 K thermal gas
using an Einstein A value of $3.471 \times 10^{-7}$ s$^{-1}$
\citep{wol98}. This estimate should be good to about a factor
of two to three,  modulo the amount of extinction, if the 
contributions from thermal and fluorescent emission are about equal. 
Table~\ref{tbl-ne2v} gives the mass of H$_2$ needed to produce the
observed line strength under the assumption of the same extinction
as used in the Pa$\alpha$ calculation.  This is a lower limit on the
H$_2$ since the extinction will be significantly higher in the dust
protected molecular region.  \citet{hou04} calculate a 9.7 \micron\ 
extinction of 0.49 magnitudes which predicts about 2.6 magnitudes 
of extinction at 2.12 \micron\ as opposed to the 1.4 magnitudes 
derived from the atomic hydrogen lines.  \citet{van00} spectroscopically 
observed a line strength that is about 2.2 times larger than that quoted 
in the table.  We may have missed some of the extended flux but the value 
here is within 1.9$\sigma$ of the spectroscopic value.   

\section{Evolution and Feedback in SBS0335-052E} \label{s-ef}

The primary conclusions on the nature of the SSCs S1 and S2 in TSKEV
remain intact.  Although the possibility exists that the diffuse emission
around S1 and S2 is due to ionizing radiation from them, the more likely
explanation is that the ionization is done by in-situ stars.  The lower
equivalent width of the diffuse radiation is strong evidence for this view.
If the diffuse emission was due to pure ionized gas with no stars its 
equivalent width should be higher rather than lower than the equivalent
width of the star rich clusters S1 and S2.  This leaves the conclusion that S1 and 
S2 are currently not producing any negative feedback from photo-ionizing
flux on star formation in SBS0335-052E intact.  The delay in photo-ionizing
feedback may also have implications for planet formation.  The ambient
UV radiation field may be significantly less than the fields calculated
without considering absorption by stellar winds, eg. \citet{fat08}. This
would increase the probability of planet formation even in the environment
of a SSC which would normally be considered not conducive to planet formation.
It also has implications on whether globular clusters are good sites for
future searches for planets.

The deeper and higher spatial resolution images have revealed that there are
star formation regions in SBS0335-052E that are not compact SSCs.  Two of 
these regions, S7 and S8, appear to be younger than S1 and S2 while the
other extended region, S3-Pa$\alpha$ appears to be roughly coeval with 
S1 and S2.  The appearance of these extended regions of star formation may 
mark the end of the formation of compact stellar clusters and perhaps
the beginning of negative photo-ionizing feedback.  Because the volume emission 
measures of the extended sources are significantly lower than that of S1 and
S2, modulo extinction, they do not contribute much to the net radio emission.
If the extended regions have extended HII regions then they should be optically
thin.  Future observations with the EVLA should be capable of taking spatially
resolved radio images of the different sources to determine whether they
are optically thick or thin at radio frequencies.

\citet{elm97} discuss the effect of pressure on the ratio of compact to diffuse
star formation.  They claim that high pressure favors the formation of compact
clusters as opposed to diffuse star formation. \cite{bek08} has simulated the
merger of two dwarf galaxies with stellar masses $4 \times 10^8 M_{\sun}$ and
gas masses of $8 \times 10^8 M_{\sun}$. $22\%$ of the new stars in this
simulation appear in compact clusters, far more than in simulations at lower
pressures.  SBS0335-052 has a gas mass of $2.1 \times 
10^9 M_{\sun}$ and a gas to stellar mass ratio of 25-26 (Pustilnik et al. 2001; this
work) therefore the central gas pressure should be significantly higher than 
the model considered by \citet{bek08} leading to an increased fraction of star 
formation in clusters.

\section{Comparison with Other Recent Work}

One day after the initial submission of this manuscript, new work on
SBS0335-052E became publicly available \citep{rei08}, hereinafter RJH, 
which reaches some different conclusions than this work.  RJH derive 
significantly lower masses and ionizing fluxes than this work, particularly 
for the youngest sources, S1 and S2, where the difference is a factor of 15
and 10 respectively.  The main diagnostic tool of RJH
is SED fitting to the optically observed HST ACS images of \citet{ost08}.
This fitting, confined to the first 4 ACS bands leads to significantly
lower dust extinction values than are derived in this work and TSKEV.
Our A$_V$ value is 12.1 as opposed to 0.5 in RJH.  This is the primary
reason for the different conclusions drawn between the two works. Source
fluxes redder than the F550M band deviate strongly from the RJH derived
SED and must be explained by a combination of hot dust and Extended Red 
Emission \citep{wit04}. The extinction model needed to account for the 
observed hydrogen emission line ratios is gray from H$\beta$ to Br$\gamma$
with 0.9 magnitudes of extinction for each line.  Our model with an A$_V$
value of 12.1 and an extinction law given by \citet{rie85} implies significantly
more line flux which requires a larger ${N_e}^2V$ and more mass in
stars to provide the ionizing flux.  This model does not require the
extra emission components to match the observed fluxes.

\section{Summary} \label{s-con}

New high spatial resolution NICMOS camera 2 images of the region of SBS0335-052E 
that contains a compact group of SSCs have improved our detailed knowledge of the
geometry and physics of the system.  There are 6 point-like objects, 7 if the binary
nature of S3 is taken into account, that are either currently SSCs (S1 and S2) or
were most likely previously SSCs (S3-S6).  We now know that the continuum source
S3 is not at the same position as the Pa$\alpha$ source S3-Pa$\alpha$.  In addition
to the compact sources there are 4 diffuse or extended regions of Pa$\alpha$ emission,
indicating current star formation in non compact objects.  It should be noted, however,
that a diffuse distribution of many compact sources with stellar masses on the order
of $10^4$ M$_{\sun}$ would be difficult to differentiate from a diffuse distribution
of individual stars.  Two of these regions, S7 and S8 appear to be significantly 
younger than the two star forming SSCs.  One of the star forming regions, S8, is 
well separated from the compact group of previously known sources.  These regions
are the first indication of significant star forming activity in SBS0335-052E in 
anything other than very compact clusters. They are significantly less massive than the
very compact clusters, modulo extinction, and may be open clusters that will not 
remain bound after the gas is expelled. The new images still support the general 
picture of delayed photo-ionizing negative feedback from the sources S1 and S2 
along with the general picture of propagating star formation from the north to 
the south. This delay in feedback may also affect the number of planetary systems
that can form in what would normally be considered a very harsh environment.  
We have also
confirmed that the H$_2$ emission is inside both S1 and S2 but is not at the 
exact center of the two objects and probably marks the location of the regions
producing the observed mid-infrared emission.

One result that differed from TSKEV is the size of S1 where the possibility
was raised that SSCs may be the origin of intermediate mass black holes through
the merging of stellar mass black holes created by supernova events in the 
cluster.  The new radius of 15 pc is approximately 3 times larger than the 
previous  radius and would affect the possibility of forming intermediate 
mass black holes. Equation 1 of \citet{zwa06} indicates that the time scale 
for forming merged intermediate mass black holes is dependent on radius to 
the 3/2 power.  The new radius of S1 would therefore increase the time scale
by about a factor of 5.  Even at the new radius,however, SSCs seem to be a 
leading candidate for the formation site of greater than stellar mass black 
holes. See also the very recent discussion of massive black hole formation
in early low metallicity star clusters by \citet{dev08}.

\acknowledgments

We wish to acknowledge very helpful discussions with Kelsey Johnson and to 
thank an anonymous referee for comments that have improved this
manuscript. This work contains data from the NASA/ESA Hubble Space Telescope 
which is
operated by the Association of Universities for Research in Astronomy
(AURA) Inc. under NASA contract NAS5-26555.

\clearpage

\begin{deluxetable}{ccccccccc}
\tabletypesize{\scriptsize}
\tablecaption{HST Image Observational Parameters \label{tbl-obs}}
\tablewidth{0pt}
\tablehead{
\colhead{Inst.} & \colhead{Prop.} & \colhead{Mode} & \colhead{Filter} &
\colhead{Samp.\tablenotemark{a}} & \colhead{NSamp.\tablenotemark{a}} &
\colhead{Int. time} & \colhead{Ints.} & \colhead{Total} \\
\colhead{} & \colhead{} & \colhead{} & \colhead{} & \colhead{} & 
\colhead{} & \colhead{sec.} & \colhead{\#} & \colhead{sec.}
}
\startdata
NICMOS & 10856 & Cam. 2 & F187N & STEP64 & 20 & 768 & 8 & 6144 \\
NICMOS & 10856 & Cam. 2 & F190N & STEP32 & 17 & 320 & 8 & 2560 \\
NICMOS & 10856 & Cam. 2 & F212N & STEP64 & 20 & 768 & 8 & 6144 \\
NICMOS & 10856 & Cam. 2 & F215N & STEP64 & 20 & 768 & 8 & 6144 \\
NICMOS & 10894 & Cam. 2 & F187N & STEP16 & 26 & 304 & 16 & 4864 \\
NICMOS & 10894 & Cam. 2 & F190N & STEP16 & 25 & 288 & 16 & 4600 \\
NICMOS & 10894 & Cam. 2 & F160W & STEP16 & 26 & 304 & 8 & 2432 \\
NICMOS & 10894 & Cam. 2 & F205W & STEP64 & 24 & 272 & 8 & 2176 \\
ACS & 10575 & WFC1 & F550M & - & - & 430 & 1 & 430 \\
ACS & 10575 & WFC1 & FR656 & - & - & 680 & 1 & 680 \\
ACS & 9470 & SBC & F140LP & - & - & 2700 & 1 & 2700 \\
\enddata

\tablenotetext{a}{NICMOS only}
\end{deluxetable}

\clearpage
\begin{deluxetable}{ccccccc}
\tabletypesize{\scriptsize}
\tablecaption{Observed Line and Continuum Measurements\tablenotemark{a} \label{tbl-flux}}
\tablewidth{0pt}
\tablehead{
\colhead{Source} & \colhead{Pa$\alpha$} & \colhead{H$_2$} & \colhead{0.14 $\mu$m} & 
\colhead{1.6 $\mu$m} & \colhead{2.05 $\mu$m} & \colhead{Pa$\alpha$ Equivalent Width} \\
\colhead{ } & \colhead{erg s$^{-1}$ cm$^{-2}$} &  \colhead{erg s$^{-1}$ cm$^{-2}$} &
 \colhead{Jy} & \colhead{Jy} & \colhead{Jy} & \colhead{\AA}
}
\startdata
S1 & $9.7 \times 10^{-15}$ & $8.8 \times 10^{-17}$ & $1.7 \times 10^{-5}$ &
$2.6 \times 10^{-5}$ & $ 4.5 \times 10^{-5}$ & 1281 \\
S2 & $7.2 \times 10^{-15}$ & $3.5 \times 10^{-17}$ & $8.6 \times 10^{-6}$ &
$2.3 \times 10^{-5}$ & $ 3.5 \times 10^{-5}$ & 1182 \\
S3-Pa$\alpha$ & $1.5 \times 10^{-15}$ & \nodata & $1.2 \times 10^{-5}$ & 
$1.9 \times 10^{-6}$ & $ 1.1 \times 10^{-5}$ & 1103 \\
S3 & \nodata & \nodata & $1.2 \times 10^{-5}$ & 
$1.1 \times 10^{-5}$ & $ 5.9 \times 10^{-6}$ & \nodata \\
S4 & \nodata & \nodata & $ 4.1 \times 10^{-5}$ & $1.2 \times 10^{-5}$ &
$ 9.2 \times 10^{-6}$ & \nodata \\
S5 & \nodata & \nodata & $ 4.9 \times 10^{-5}$ & $2.4 \times 10^{-5}$ &
$ 1.9 \times 10^{-5}$ & \nodata \\
S6 & \nodata & \nodata & $ 4.1 \times 10^{-6}$ & $4.1 \times 10^{-6}$ &
$ 3.4 \times 10^{-6}$ & \nodata \\
S7  & $1.1 \times 10^{-15}$ & \nodata & $6.0 \times 10^{-7}$ & $3.7 \times 10^{-6}$ &
$ 4.3 \times 10^{-6}$ & 3478\\
S8 & $3.4 \times 10^{-16}$ & \nodata & $ 1.2 \times 10^{-6}$ & $1.2 \times 10^{-6}$ &
$ 1.6 \times 10^{-6}$ & 2079 \\
diffuse & $1.37 \times 10^{-14}$ & \nodata & $ 5.7 \times 10^{-4}$ & $2.6 \times 10^{-5}$ &
$ 2.1 \times 10^{-5}$ & 672 \\
\enddata

\tablenotetext{a}{All fluxes are for an aperture diameter of 5 pixels ($0.378\arcsec$) 
and the equivalent widths are for an aperture diameter of 6 pixels ($0.454\arcsec$).
Flux and equivalent width errors are estimated as $10\%$ and are dominated by the 
amount of extended flux.}
\end{deluxetable}

\clearpage

\begin{deluxetable}{cccccccc}
\tabletypesize{\scriptsize}
\tablecaption{Line Power, ${N_e}^2V$, and Star Formation Rates \label{tbl-ne2v}}
\tablewidth{0pt}
\tablehead{
\colhead{Source} & \colhead{Pa$\alpha$ Line Power\tablenotemark{a}} & 
\colhead{${N_e}^2V$\tablenotemark{b}} & 
\colhead{Ext. Cor. ${N_e}^2V$\tablenotemark{c}} & \colhead{SFR}
& \colhead{Ext. Cor. H$_2$ Pow.} & \colhead{Ext. Cor. H$_2$ Mass} & 
\colhead{Stellar Mass\tablenotemark{d}}\\
\colhead{ } & \colhead{ergs/sec} & \colhead{cm$^{-3}$} & \colhead{cm$^{-3}$} &
\colhead{M$_{\sun}$ yr$^{-1}$} & \colhead{ergs/sec} & \colhead{M$_{\sun}$} & 
\colhead{M$_{\sun}$}
}
\startdata
S1 & $3.4 \times 10^{39}$ & $1.8 \times 10^{65}$ & $7.9 \times 10^{65}$ & 1.2
& $1.1 \times 10^{38}$ & 21. & $1.5 \times 10^7$\\
S2 & $2.5 \times 10^{39}$ & $1.3 \times 10^{65}$ & $5.9 \times 10^{65}$ & 0.9
& $4.5 \times 10^{37}$ & 8.4. & $1.1 \times 10^7$\\
S3-Pa$\alpha$ & $5.3 \times 10^{38}$ & $2.7 \times 10^{64}$ & $1.2 \times 10^{65}$ & 0.18
& \nodata & \nodata & $2.3 \times 10^6$\\
S7 & $3.9 \times 10^{38}$ & $2.0 \times 10^{64}$ & $9.0 \times 10^{64}$ & 0.14
& \nodata & \nodata & $9.4 \times 10^4$\\
S8 & $1.2 \times 10^{38}$ & $6.2 \times 10^{63}$ & $2.8 \times 10^{64}$ & 0.043
& \nodata & \nodata & $2.2 \times 10^5$\\
diffuse & $4.9 \times 10^{39}$ & $2.5 \times 10^{65}$ & $5.5 \times 10^{65}$ & 0.85 
& \nodata & \nodata & $2.05 \times 10^7$\\
\enddata
\tablenotetext{a}{Assumes a distance of 54.3 Mpc and no correction for extinction}
\tablenotetext{b}{No correction for extinction}
\tablenotetext{c}{Assumes Pa$\alpha$ extinction by a factor of 4.5 (A$_V$ = 12.1 
for all sources except for the diffuse emission where it is a factor of 2.25 
(see \S~\ref{ss-s1s2}).}
\tablenotetext{d}{See \S~\ref{s-mass} for the calculation of the stellar mass.}
\end{deluxetable}

\clearpage

\clearpage

\clearpage

\clearpage
\begin{figure}
\plotone{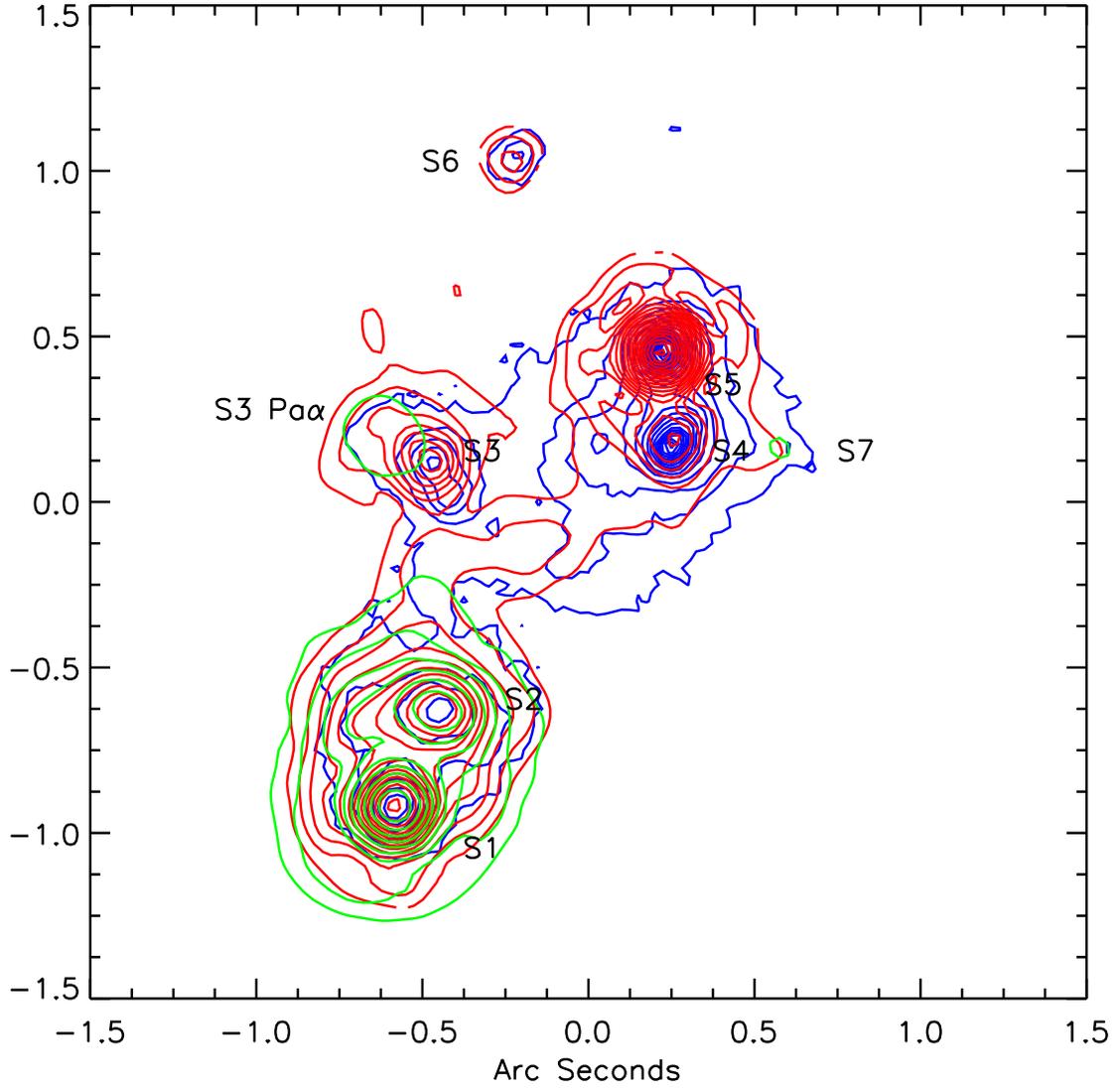}
\caption{Combined contours of the ACS UV emission (blue), Near Infrared continuum (red)
and Pa$\alpha$ (green).  Note the relative offsets of the UV continuum, Near
Infrared continuum and Pa$\alpha$ for the source S3. North is up and East is to
the left.}
\label{fig-comb}
\end{figure}

\clearpage

\begin{figure}
\plotone{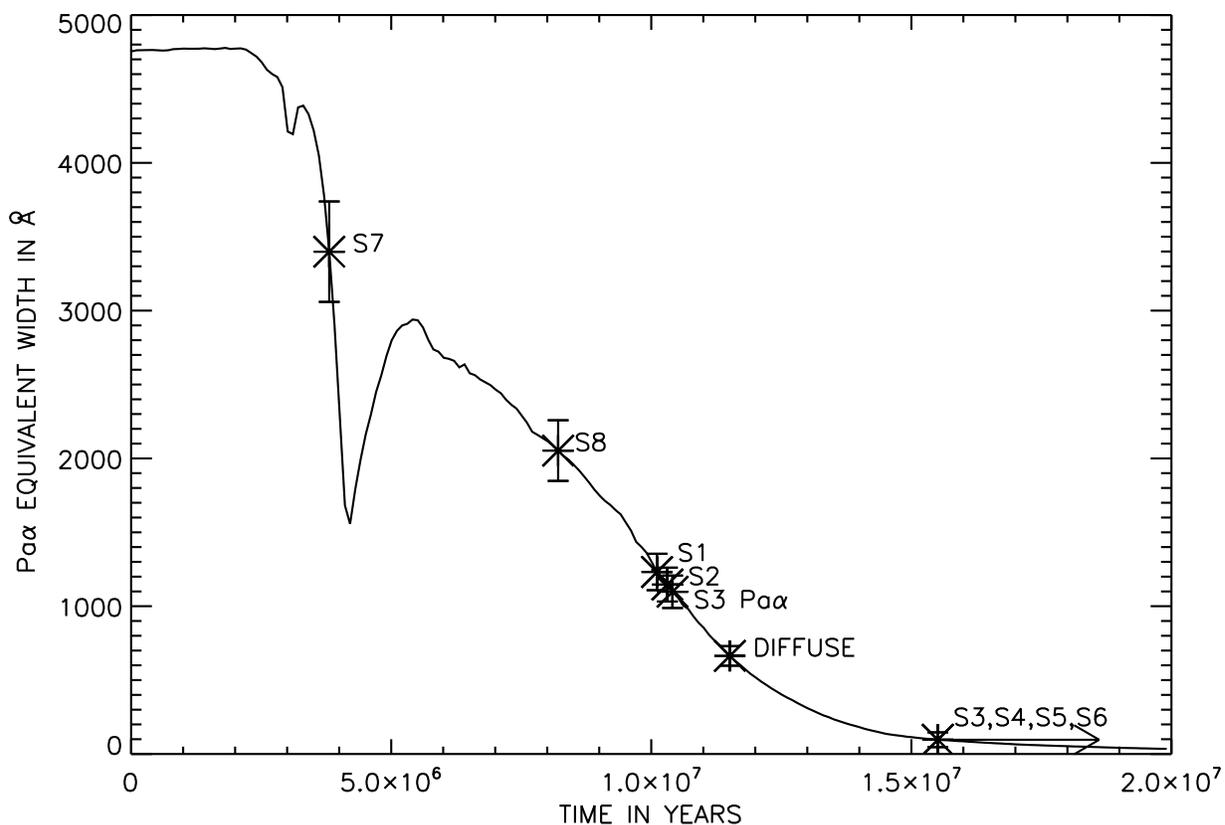}
\caption{The solid line represents the equivalent width of Pa$\alpha$ versus
time calculated from a Starburst99 simulation.  The asterisks are placed on
this line at the location of the equivalent widths measured for each source.
The location of the position of the S8 equivalent width is at the longest time
since the simulated galaxy spends very little time at the same value of the
equivalent width that is achieved at two earlier times. The dip in equivalent
width at about $4 \times 10^6$ years is due to an increase in continuum flux
rather than a decrease in line emission.  It marks the brightest continuum
epoch of the cluster.}
\label{fig-eqw}
\end{figure}

\clearpage

\begin{figure}
\plotone{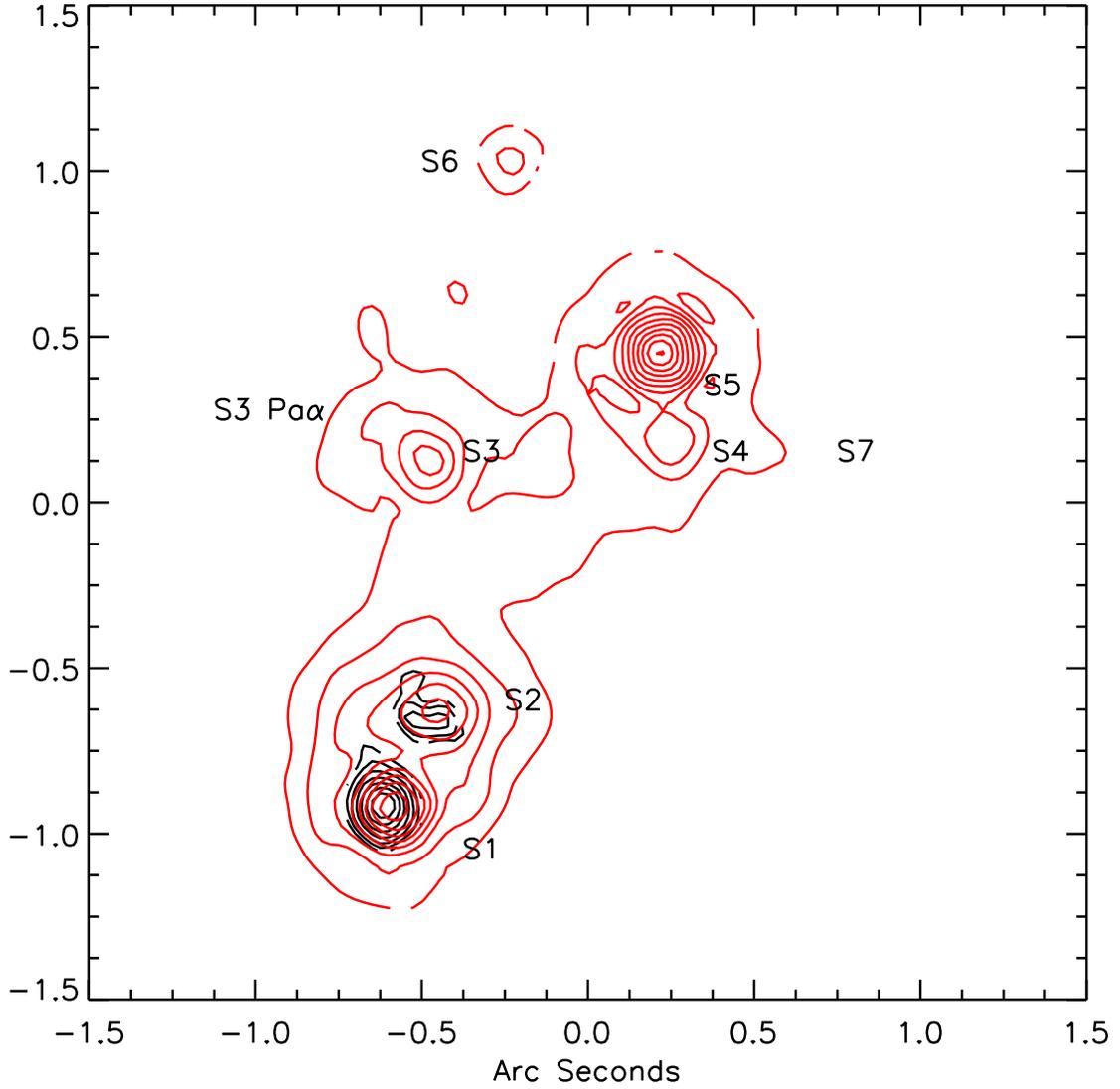}
\caption{The near infrared continuum contours (red) and H$_2$ emission (black).
Note the slight offset of the H$_2$ emission to the east of the near
infrared continuum for S1 and to the south for S2. North is up and East is to
the left.}
\label{fig-h2}
\end{figure}

\end{document}